\documentclass[11pt,twoside]{article} 
\usepackage{acta-info}
\usepackage{euler}

\newcommand{\vol}{\textbf{2,} 2 (2010) 154--167}   
\newcommand{\amss}{\amssubj{%
62N03
}}     
\newcommand{\CRs}{\CRsubj{%
D.2.5
}}   
\newcommand{\keyw}{\keywords{%
testing, C++, template metaprogramming
}}   
\begin{document}

\title{%
Testing by C++ template metaprograms
}
\maketitle

\oneauthor{
Norbert Pataki
}{
\href{http://plcportal.inf.elte.hu/en/Pages/default.aspx}{Dept. of Programming Languages and Compilers}\\
\href{http://www.inf.elte.hu/english/Lapok/default.aspx}{Faculty of Informatics}, 
\href{http://www.elte.hu/en}{E\"otv\"os Lor\'and University}\\ 
P\'azm\'any P\'eter s\'et\'any 1/C H-1117 Budapest, Hungary
}{
\href{mailto:patakino@elte.hu}{patakino@elte.hu}
}


\short{
N. Pataki
}{
Testing by C++ template metaprograms
}

\begin{abstract}
Testing is one of the most indispensable tasks in software engineering. 
The role of testing in software development has grown significantly
because testing is able to reveal defects in the code in an early stage of 
development. Many unit test frameworks compatible with C/C++ code exist, 
but a standard one is missing. Unfortunately, many unsolved problems can 
be mentioned with the existing methods, for example usually external tools 
are necessary for testing C++ programs.

In this paper we present a new approach for testing C++ programs. Our solution
is based on C++ template metaprogramming facilities, so it can work with the
standard-compliant compilers. The metaprogramming approach ensures that the
overhead of testing is minimal at runtime. This approach also supports that
the specification language can be customized among other advantages. 
Nevertheless, the only necessary tool is the compiler itself.
\end{abstract}

\section{Introduction}
\label{intro}

Testing is the most important method to check programs' correct behaviour.
Testing can reveal many problems within the code in development phase.
Testing is cruicial from the view of software quality \cite{biczo:testgen}. Many
purposes of testing can be, for instance, quality assurance, verification
and validation, or reliability estimation. Nonetheless, testing is 
potentially endless. It can never completely identify all the defects within the 
software. The main task is is to deliver faultless software \cite{szabo:observations}.

Correctness testing and reliability testing are two major areas of testing.
However, many different testing levels are used. In this paper we deal with
unit tests that is about correctness. 
The goal of unit testing is to isolate each part of the program and to show
that the individual parts are correct. A unit test provides a strict, 
written contract that the piece of code must satisfy. As a result, it 
affords several benefits. Unit tests find problems early in the development phase.
Unfortunately, most frameworks need external tools \cite{Hamill:frameworks}.

A testing framework is proposed in \cite{bagge:testing,bagge:axiom} which is based
on the C++0x -- the C++ forthcoming standard. The framework takes advantage of
\emph{concepts} and \emph{axioms}. These constructs support the generic programming
in C++ as they enable to write type constraints in template parameters. By now,
these constructs are removed from the draft of the next standard.
Metaprogram testing framework has already been developed \cite{sinkovics:unittest}
too, but it deals with metaprogams, it is just the opposite of our approach.

C++ template metaprogramming is an emerging paradigm which enables
to execute algorithms when ordinary C++ programs are compiled.
The style of C++ template metaprograms is very similar to the
functional programming paradigm. Metaprograms have many advantages
that we can harness. Metalevel often subserves the validation \cite{devai:meta}.

Template metaprograms run at compilation-time, whereupon the overhead
at runtime is minimal. Metaprograms' ``input'' is the runtime C++ program
itself, therefore metaprograms are able to retrieve information about the
hosting program. This way we can check many properties about the programs
during compilation
\cite{mihalicza:access,gsd:debug,szugyi:methods,szugyi:sophisticatedmethods}.

Another important feature of template metaprograms is the opportunity of
\emph{domain-specific languages}. These special purpose languages are
integrated into C++ by template metaprograms \cite{czarnecki:dsl,gil:sql}.
Libraries can be found that support the development of domain-specific languages
\cite{karlsson:boostbook}. New languages can be figured out to write C++
template metaprograms \cite{sipos:eclean}. Special specification languages
can be used for testing C++ programs without external tools.

In this paper we present a new approach to test C++ code. Our framework is
based on the metaprogramming facility of C++. We argue for testing by
meta-level because of numerous reasons.

The rest of this paper is organized as follows. In Section \ref{TMP}
C++ template metaprograms are detailed. In Section \ref{framework}
we present the basic ideas behind our approach, after that in
Section \ref{eval} we analyze the advantages and disadvantages of
our framework. Finally, 
the future work is
detailed in Section \ref{future}.

\section{C++ template metaprogramming}
\label{TMP}

The template facility of C++ allows writing algorithms and data structures
parametrized by types. This abstraction is useful for designing general
algorithms like finding an element in a list. The operations of lists of
integers, characters or even user defined classes are essentially the same.
The only difference between them is the stored type. With templates we can 
parametrize these list operations by type, thus, we have to write the 
abstract algorithm only once. The compiler will generate the integer, 
double, character or user defined class version of the list from it. 
See the example below:

\begin{verbatim}
template<typename T>
struct list 
{
  void insert( const T& t );
  // ...
};

int main() 
{
  list<int> l;      //instantiation for int
  list<double> d;   // and for double
  l.insert( 42 );   // usage
  d.insert( 3.14 ); // usage
}
\end{verbatim}

The list type has one template argument \texttt{T}. This refers to the parameter
type, whose objects will be contained in the list. To use this list we have to 
generate an instance assigning a specific type to it. The process is called 
\emph{instantiation}. During this process the compiler replaces the abstract 
type \texttt{T} with a specific type and compiles this newly generated code. 
The instantiation can be invoked explicitly by the programmer but in
most cases it is done implicitly by the compiler when the new list is first 
referred to.

The template mechanism of C++ enables the definition of partial and full 
\emph{specializations}. Let us suppose that we would like to create a more 
space efficient type-specific implementation of the \texttt{list} template 
for the \texttt{bool} type. We may define the following specialization:

\begin{verbatim}
template<>
struct list<bool> 
{
  //type-specific implementation
};
\end{verbatim}

The implementation of the specialized version can be totally different from 
the original one. Only the names of these template types are the same. If 
during the instantiation the concrete type argument is \texttt{bool}, the 
specific version of \texttt{list<bool>} is chosen, otherwise the general one 
is selected.

Template specialization is an essential practice for template metaprogramming
too \cite{cpptmp}. In template metaprograms templates usually refer to other
templates, sometimes from the same class with different type argument. In this
situation an implicit instantiation will be performed. Such chains of recursive
instantiations can be terminated by a template specialization. See the
following example of calculating the factorial value of 5:

\begin{verbatim}
template<int N>
struct Factorial 
{
  enum { value=N*Factorial<N-1>::value };
};

template<>
struct Factorial<0> 
{
  enum { value = 1 };
};

int main() 
{
  int result = Factorial<5>::value;
}
\end{verbatim}

To initialize the variable \texttt{result} here, the expression 
\texttt{Factorial<5>::value} has to be evaluated. As the template argument is 
not zero, the compiler instantiates the general version of the 
\texttt{Factorial} template with 5. The definition of \texttt{value} is 
\texttt{N * Factorial<N-1>::value}, hence the compiler has to instantiate 
\texttt{Factorial} again with 4. This chain continues until the concrete 
value becomes 0. Then, the compiler chooses the special version of 
\texttt{Factorial} where the \texttt{value} is 1. Thus, the instantiation chain 
is stopped and the factorial of 5 is calculated and used as initial value
of the \texttt{result} variable in \texttt{main}. This metaprogram ``runs" 
while the compiler compiles the code. 

Template metaprograms therefore stand for the collection of templates, their 
instantiations and specializations, and perform operations at compilation 
time. The basic control structures like iteration and condition appear in 
them in a functional way \cite{shp:metafun}. As we can see in the previous 
example iterations in metaprograms are applied by recursion. Besides, the 
condition is implemented by a template structure and its specialization.

\begin{verbatim}
template<bool cond,class Then,class Else>
struct If
{
  typedef Then type;
};

template<class Then, class Else>
struct If<false, Then, Else>
{
  typedef Else type;
};
\end{verbatim}

The \texttt{If} structure has three template arguments: a boolean and two
abstract types. If the \texttt{cond} is false, then the partly-specialized
version of \texttt{If} will be instantiated, thus the \texttt{type} will be
bound to \texttt{Else}. Otherwise the general version of \texttt{If}
will be instantiated and \texttt{type} will be bound to \texttt{Then}.

With the help of \texttt{If} we can delegate type-related decisions from
design time to instantiation (compilation) time. Let us suppose, we want
to implement a \texttt{max(T,S)} function template comparing values of
type T and type S returning the greater value. The problem is how 
we should define the return value. Which type is ``better'' to return
the result? At design time we do not know the actual type of the \texttt{T}
and \texttt{S} template parameters. However, with a small template metaprogram
we can solve the problem:

\begin{verbatim}
template <class T, class S>
typename If<sizeof(T)<sizeof(S),S,T>::type
  max( T x, S y)
  {
    return x > y ? x : y;
  }
\end{verbatim}

Complex data structures are also available for metaprograms. Recursive
templates store information in various forms, most frequently as tree
structures, or sequences. Tree structures are the favorite forms of
implementation of expression templates \cite{veldhuizen:expression}.
The canonical examples for sequential data structures are \texttt{typelist}
\cite{alexandrescu:modern} and the elements of the \texttt{boost::mpl}
library \cite{karlsson:boostbook}.

We define a typelist with the following recursive template:

\begin{verbatim}
class NullType {};

typedef Typelist<char,Typelist<signed char,
    Typelist<unsigned char,NullType> > >
  Charlist;
\end{verbatim}

In the example we store the three character types in a typelist.
We can use helper macro definitions to make the syntax more readable.

\begin{verbatim}
#define TYPELIST_1(x)           
          Typelist< x, NullType>
#define TYPELIST_2(x, y)        
          Typelist< x, TYPELIST_1(y)>
#define TYPELIST_3(x, y, z)     
          Typelist< x, TYPELIST_2(y,z)>
// ...
typedef 
TYPELIST_3(char,signed char,unsigned char)  
  Charlist;
\end{verbatim}

Essential helper functions -- like \texttt{Length}, which computes the size 
of a list at compilation time -- have been defined in Alexandrescu's Loki
library \cite{alexandrescu:modern} in pure functional programming style.
Similar data structures and algorithms can be found in the
metaprogramming library \cite{karlsson:boostbook}.

The examples presented in this section expose the different approaches of
template metaprograms and ordinary runtime programs. Variables are
represented by static constants and enumeration values, control structures
are implemented via template specializations, functions are replaced by 
classes. We use recursive types instead of the usual data structures. 
Fine visualizer tools can help a lot to comprehend these
structures \cite{visualization}.

\section{Testing framework}
\label{framework}

In this section we present the main ideas behind our testing
framework which takes advantage of the C++ template metaprogramming.

First, we write a simple type which makes connection between the
compilation-time and the runtime data. This is the kernel of the
testing framework. If the compilation-time data is not equal to
the runtime data, we throw an exception to mark the problem.

\begin{verbatim}
struct _Invalid
{
  // ...
};

template < int N >
class _Test
{
  const int value;

public:

  _Test( int i ) : value( i )
  {
    if ( value!=N )
      throw _Invalid();
  }

  int get_value() const
  {
    return value;
  }
};
\end{verbatim}

Let us consider that a runtime function is written, that calculates the 
factorial of its argument. This function is written in an iterative way:

\begin{verbatim}
int factorial( int n )
{
  int f = 1;
  for( int i = 1; i <= n; ++i)
  {
    f *= i;
  }
  return f;
}
\end{verbatim}

It is easy to test the factorial function:

\begin{verbatim}
template <int N>
_Test<Factorial<N>::value> factorial_test( const _Test<N>& n )
{
  return factorial( n.get_value() );
}
\end{verbatim}

When \texttt{factorial\_test} is called, it takes a compile-time and runtime parameter.
The constructor of \texttt{\_Test} guarantees, that the two parameters are equal. We
take advantage of the parameter conversions of C++. When an integer is passed as
\texttt{\_Test}, it automatically calls the constructor of \texttt{\_Test} which tests
if the runtime and compilation time parameters are the same. If the
runtime and compilation time parameters disagree, an exception is raised. The return
type of \texttt{factorial\_test} describes that it must compute the \texttt{Factorial<N>}.
When it returns a value, it also calls the constructor of \texttt{\_Test}.
At compilation time it is computed what the return should be according to the
metaprogram specification -- e.g. what the \texttt{Factorial<N>} is.
Because the \texttt{factorial\_test} takes a \texttt{\_Test} parameter, two parameters
cannot be different. When the \texttt{factorial\_test} returns it is also evaluates
if the result of compilation time algorithm is the same with the result of the runtime
algorithm, and an exception raised if it fails. So, we have a runtime and compilation
time input, first we calculate the result at compilation time from the compilation
time input. At runtime we have the very same input and a runtime function, and evaluates
if the runtime algorithm results in the very same output. If it fails an exception is
thrown.

Of course, we have to call the \texttt{factorial\_test} function:

\begin{verbatim}
int main()
{
  factorial_test< 6 >( 6 );
}
\end{verbatim}

In this case, we write \texttt{Factorial} metafunction
that counts the factorial at compilation time, but we do not
have to write this metafunction with metaprograms. This metaprogram can
be generated by the compiler from a specification that can be
defined in EClean \cite{sipos:eclean, shp:metafun}, Haskell, or other 
domain-specific language \cite{gsd:dslparser}.

Instead of return value, references are often used to transmit data to the caller:

\begin{verbatim}
void inc( int& i )
{
  ++i;
}
\end{verbatim}

At this point, we cannot wrap the call of this function into a tester function.
Hence, in this case we deal with a new local variable to test.

\begin{verbatim}
template < int N >
_Test<N+1> inc_test( const _Test<N>& n )
{
  int i = n.get_value();
  inc( i );
  return i;
}
\end{verbatim}

Since doubles cannot be template arguments we have to map doubles to
integers. The natural way to do this mapping is the usage of the
significand and exponent. Here is an example, that presents this idea:

\begin{verbatim}
template <int A, int B>
struct MetaDouble
{
};


double f( double d )
{
  return d*10;
}

template < int A, int B >
class _Test
{
  const double d;
public:
  _Test( double x ): d( x )
  {
    if ( get_value() != d )
      throw _Invalid();
  }

  double get_value() const
  {
    return A * pow( 10.0L, B );
  }
};

template <int A, int B>
_Test<A,B+1> f_test(MetaDouble<A, B> d)
{
  double dt = A*pow( 10.0L, B );
  return f( dt );
}
\end{verbatim}

This framework can be easily extended in the way of C++ Standard Template Library (STL)
\cite{stroustrup:cpp}. We may use functor objects instead of the equality operator to
make the framework more flexible because it can test more relations. We can take
advantage of default template parameters of template classes. The following code 
snippet can be applied to the integers:

\begin{verbatim}
template <int N, class relation = std::equal_to<int> >
class _Test
{
  const int value;
  const relation rel;

public:
  _Test( int i ) : value( i )
  {
    if ( rel(N, value) )
      throw _Invalid();
  }

  int get_value() const
  {
    return value;
  }
};
\end{verbatim}

\section{Evaluation}
\label{eval}

In this section we argue for our approach. We describe pros and cons
and present scenarios where our method is more powerful then the
existing ones.

One the most fundamental advantages is that our framework does not need
external tools, the only necessary tool is the compiler itself. Nevertheless,
another important feature, that we compute the result at compilation time, 
so the runtime overhead is minimal. Of course, the compilation time is
increased. The performance analysis of C++ template metaprograms is
detailed in \cite{gsd:profiling}.

Our approach is able to detect and pursue the changes external APIs' interface.
For instance, the type of return value has been changed, we do not need to
modify the specifications. Just like the \texttt{max} example in \ref{TMP}
section, metaprograms can determine the type of return values, etc.

Domain-specific languages can be developed with the assistance of template metaprograms.
Therefore, specification languages can be easily adopted to our approach. Users
can select a specification language from the exisiting ones or develop new 
domain-specific languages for the specification \cite{wgt:dsl}. The usual 
specification methods support only one specification language at all.

Moreover, metaprograms are written in an functional way, but runtime C++ programs
are written in an imperative way. Therefore, testing approach and implementation
is quite different. It is easy to focus on the results this way.
A very useful advantage is that that our framework can be used for legacy
code too.

Albeit there are some typical approaches which cannot be tested with our method.
For instance, metaprograms cannot access database servers and
metaprograms cannot deal with other runtime inputs. Files and requests
cannot be managed with metaprograms. On the other hand, we can test
the business logic of the programs: is the result correct if the input
would be the specificated one. Also, calls of virtual methods cannot
be handled at compilation time.

Our approach cannot facilitate the testing of multithreaded programs.
Testing concurrent programs is hard, but the compiler acts as a
single-threaded non-deterministic interpreter.

\section{Conclusions and future work}
\label{future}

Testing is one of the most important methods to ensures programs' correctness.
In this paper we argue for a new approach to test C++ programs. Our solution
takes advantage of C++ template metaprogramming techniques in many ways.
We have examined the pros and cons of our method.

After all, the most important task is to work out a set of special
specification languages and generate standard compliant C++ metaprograms
from these specifications.

In this paper we argue for a method that manages runtime entities at compilation
time. With this method we tested runtime functions. Many other interesting
properties should be managed in this way, for instance, the runtime complexity or
the prospective exceptions.

Another important task is developing mapping between the runtime and compile time
advanced datastructures. Node-based datastructures (like trees, linked lists) are
also available in metalevel, but we have not mapped these structures to runtime
akins. User-defined classes also may be mapped to the their compilation-time
counterparts.

An other opportunity is that we take advantage of the metalevel and
generate testcases at compilation time. In our approach the users specificate
the test cases. It would be more convenient if the compiler could generate testcases
which covers most of execution paths.

\bigskip
\rightline{\emph{Received:  August 5,  2010 {\tiny \raisebox{2pt}{$\bullet$\!}} Revised: October 15, 2010}} 

\end{document}